\documentclass[article, nojss]{jss}
\usepackage{algorithm2e}
\usepackage{mdframed}
\usepackage{amsmath,amssymb}
\usepackage{graphicx}
\usepackage{hyperref}
\newcommand{\boldX}{\pmb X}
\newcommand{\boldZ}{\pmb Z}
\newcommand{\boldLambda}{\boldsymbol \lambda}
\newcommand{\boldBeta}{\boldsymbol \beta}
\newcommand{\mR}{\mathcal{R}}
\usepackage{subfig}
\usepackage{tikz}
\usepackage{xcolor}

\usetikzlibrary{trees}

\title{Software for Distributed Computation on Medical Databases: A
  Demonstration Project}

\author{B.~Narasimhan \\ Stanford University \And D.~L.~Rubin \\
  Stanford University \AND S.~M.~Gross \\ Stanford University \And
  M.~Bendersky \\ Stanford University \And P.~W.~Lavori \\ Stanford
  University }

\Plainauthor{B. Narasimhan, D. L. Rubin, S. M. Gross, M. Bendersky,
  P. W. Lavori}

\Plaintitle{Software for Distributed Computation on Medical Databases:
  A Demonstration Project}

\Abstract{Bringing together the information latent in distributed
  medical databases promises to personalize medical care by enabling
  reliable, stable modeling of outcomes with rich feature sets
  (including patient characteristics and treatments received).
  However, there are barriers to aggregation of medical data, due to
  lack of standardization of ontologies, privacy concerns, proprietary
  attitudes toward data, and a reluctance to give up control over end
  use. Aggregation of data is not always necessary for model
  fitting. In models based on maximizing a likelihood, the
  computations can be distributed, with aggregation limited to the
  intermediate results of calculations on local data, rather than raw
  data. Distributed fitting is also possible for singular value
  decomposition. There has been work on the technical aspects of
  shared computation for particular applications, but little has been
  published on the software needed to support the ``social
  networking'' aspect of shared computing, to reduce the barriers to
  collaboration. We describe a set of software tools that allow the
  rapid assembly of a collaborative computational project, based on
  the flexible and extensible \proglang{R} statistical software and
  other open source packages, that can work across a heterogeneous
  collection of database environments, with full transparency to allow
  local officials concerned with privacy protections to validate the
  safety of the method. We describe the principles, architecture, and
  successful test results for the site-stratified Cox model and
  rank-$k$ Singular Value Decomposition. }


\Plainkeywords{distributed computation, cox regression, singular value
  decomposition, data aggregation, R}

\Address{
    Balasubramanian Narasimhan\\
    Department of Health Research and Policy, and\\
    Department of Statistics\\
    Stanford University\\
    Stanford, CA 94305\\
    E-mail: \email{naras@stat.stanford.edu}\\
    URL: \url{http://statweb.stanford.edu/~naras/}\\

    Daniel L.~Rubin\\
    Department of Radiology\\
    Stanford University\\
    Stanford, CA 94305\\
    E-mail: \email{dlrubin@stanford.edu}\\
    URL: \url{http://stanford.edu/~rubin/}\\

    Samuel M.~Gross\\
    Department of Statistics\\
    Stanford University\\
    Stanford, CA 94305\\

    Marina Bendersky\\
    Department of Radiology\\
    Stanford University\\
    Stanford, CA 94305\\

    Philip W.~Lavori\\
    Department of Health Research and Policy\\
    Stanford University\\
    Stanford, CA 94305\\
    E-mail: \email{lavori@stanford.edu}\\
    URL: \url{https://med.stanford.edu/profiles/philip-lavori}
}

\begin{document}


\section{Introduction}
\label{sec:intro}

Bringing together the information latent in distributed medical
databases promises to personalize medical care by enabling reliable,
stable modeling of outcomes with rich feature sets (including patient
characteristics and treatments received). To realize that promise, the
National Institutes of Health (NIH) and other organizations support
the aggregation of data from such databases, particularly the data
developed by investigators on NIH-supported projects. These aggregated
databases are made available to qualified investigators to explore and
extract the useful information they contain. However, there are high
(and growing) barriers to aggregation of medical data, some of them
having to do with lack of standardization of ontologies, others owing
to privacy concerns, and still others arising from a generally
proprietary attitude toward one's own institution's data, and a
reluctance to give up control. In addition, the sheer size and
complexity of some databases has caused the NIH to think about
splitting the storage of aggregated databases across several centers;
see for example \citet{nih_collab}.

It has long been known that aggregation of data is not always
necessary for model fitting. In many circumstances, such as in fitting
models based on maximizing a likelihood, the computations can be
distributed, with aggregation limited to the intermediate results of
calculations on local data, rather than raw data; see
\citet{murtagh_securing_2012} for example. Indeed, sometimes
distribution of the calculation among sites is necessary to share a
heavy computational burden, as would be the case for fitting the ADMM
models of \citet{admm}. It can even help attack the ``conflicting
ontologies'' problem, by shifting the task of translation to the
sites.

There has been work on the technical aspects of shared computation for
particular applications: \citet{jiang_webglore:_2013} and
\citet{Luocv083} describe distributed implementations of logistic
regression and Cox regression respectively without sharing
patient-level data using \proglang{Java}
applets; \citet{wolfson2010datashield} describe fitting generalized
linear models using \proglang{R}. Little has been published on the
software needed to support the ``social networking'' aspect of shared
computing, to reduce the barriers to collaboration. We present a
general set of software tools that allow the rapid assembly of a
collaborative computational project, based on the flexible and
extensible \proglang~\citep{r-core} statistical software and other
open source packages, that can work across a heterogeneous collection
of database environments (RedCAP, i2b2, local versions), with full
transparency to allow local officials concerned with privacy
protections to validate the safety of the method. Our work differs
from the above in that it is far more general: Generalized Linear
Models, Cox Models, Singular Value Decomposition computations all fit
under the same infrastructure. It also has potential for further
extensibility since \proglang{R} already has a vast array of
statistical computations built into it; distributed implementations
can be easily constructed by re-engineering existing code for those
computations. \citet{wolfson2010datashield} comes closest in spirit to
what we propose although the implementation is in a specific context,
that of the OPAL system.  Our software, examples, and documentation
can be found on CRAN as well as GitHub, and it is freely modifiable by
users. The rationale, scientific uses, and further details of the
social networking aspects of the method are discussed elsewhere. In
this article we describe the principles, architecture, and successful
test results for two privacy-preserving examples. The first is a novel
instance of fitting a site-stratified Cox Proportional Hazards Model
and the second is a well-known distributed version of Rank-$k$
Singular Value Decomposition. These illustrate the challenges and
solutions to implementing a shared computation. We show how to take
any model fit that breaks up in a similar way and implement it as a
distributed computation. We describe some possible use cases that may
be of interest.

\begin{figure}
  \centering \subfloat[\label{fig:defineComputation} Defining the
  computation. A computation may be define on any machine where
  \proglang{R} and the \pkg{distcomp} package are installed. The
  function \code{defineNewComputation()} launches a \pkg{shiny} app
  that leads the user through the process. The end result is an
  \proglang{R} data file unambiguously defining the computation
  instance for \pkg{distcomp}] {\includegraphics[trim=0in 7in 0in 0in,
    width=0.8\textwidth]{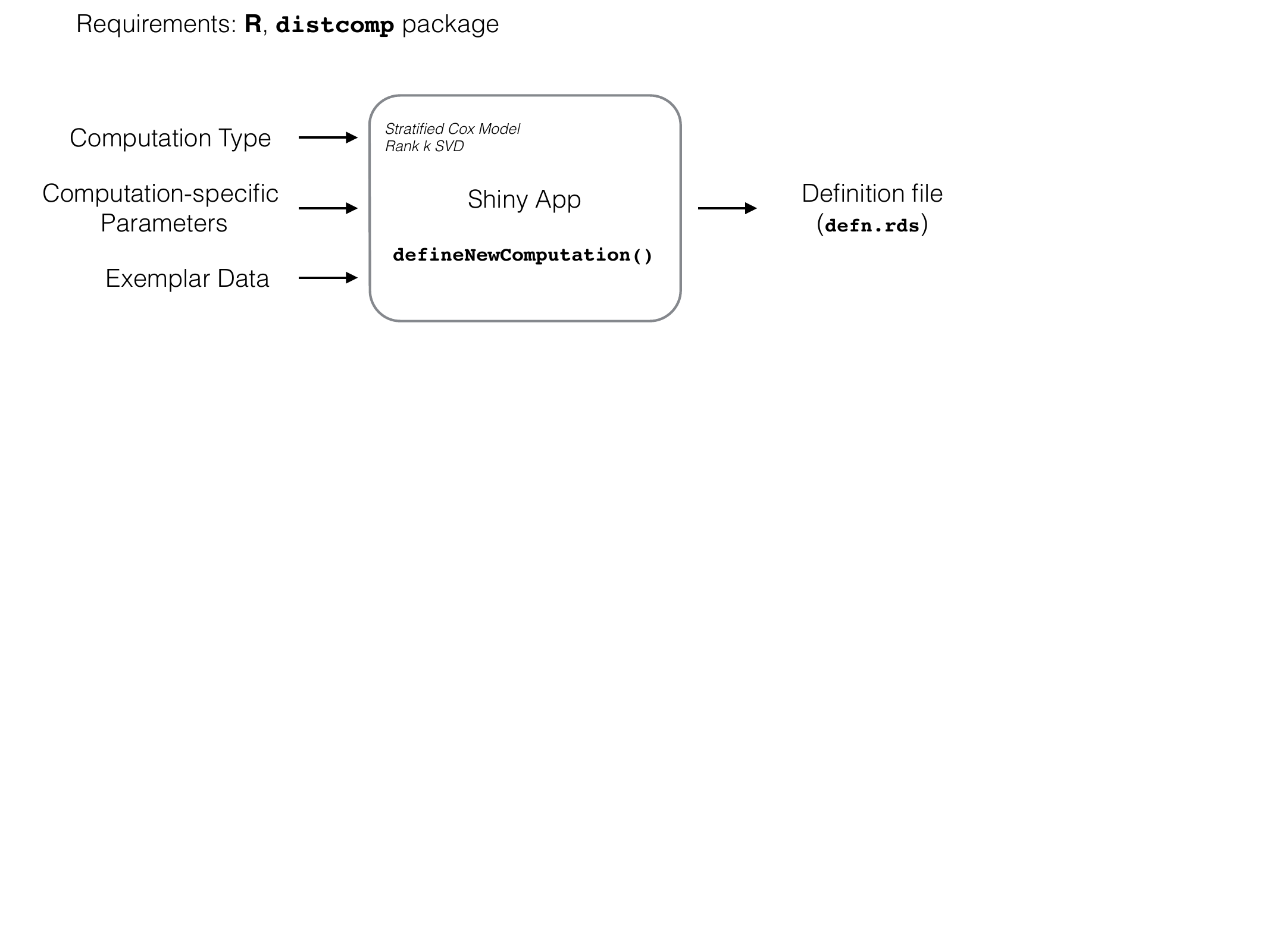}}

  \subfloat[\label{fig:setupWorker}Setting up the worker. This requires
  a one-time configuration of an \pkg{opencpu} server with the \proglang{R}
  package \pkg{distcomp} and a writable workspace. All interaction
  is through a \pkg{shiny} app that configures the worker for the
  computation.]  {\includegraphics[trim=0in 7in 0in 0in,
    width=0.8\textwidth]{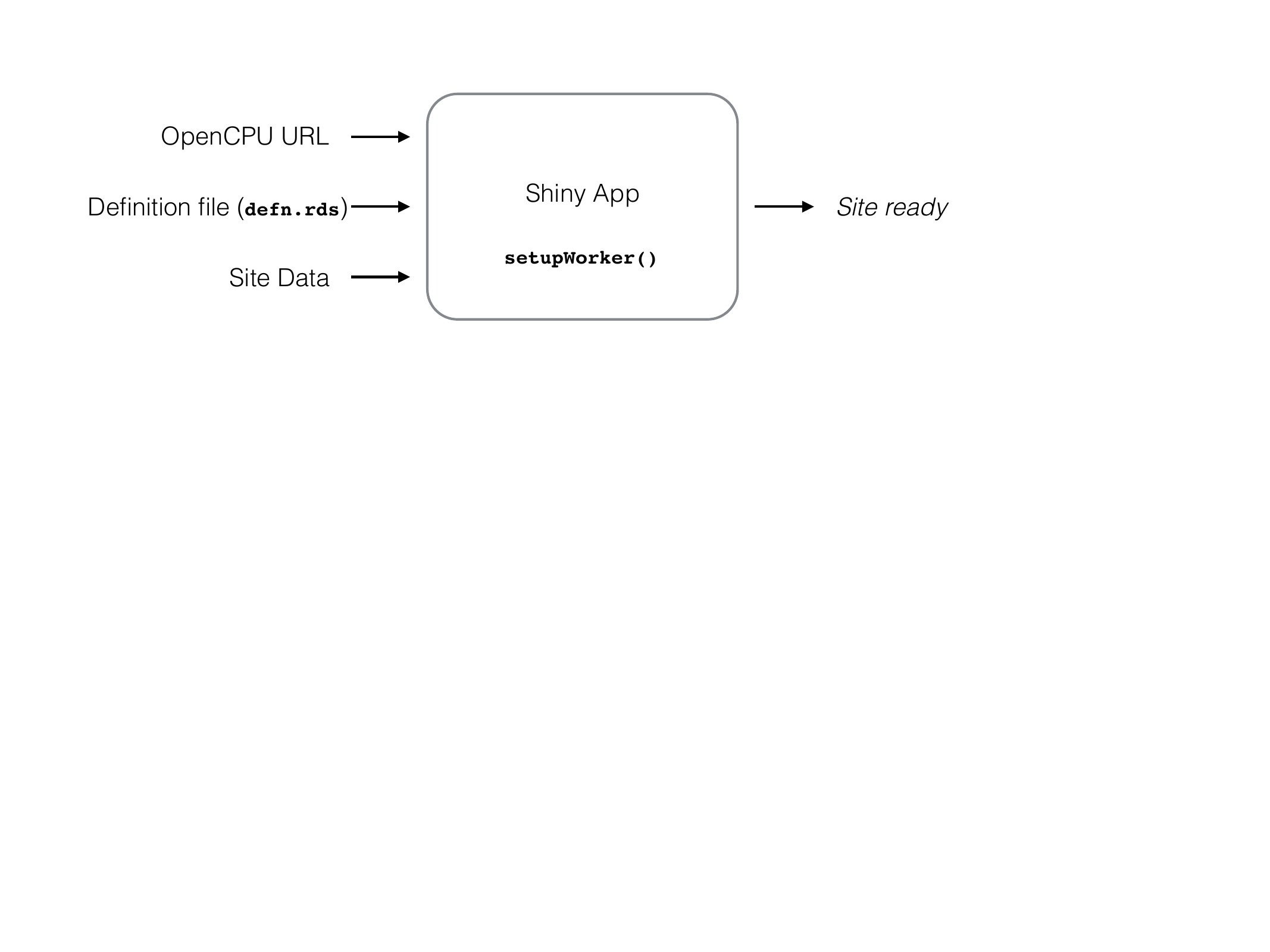}}

  \subfloat[\label{fig:setupMaster}Setting up the master. The
  \pkg{shiny} app accepts the URLs of all worker sites, checks them
  and writes an \proglang{R} script for the computation.]
  {\includegraphics[trim=0in 7in 0in 0in,
    width=0.8\textwidth]{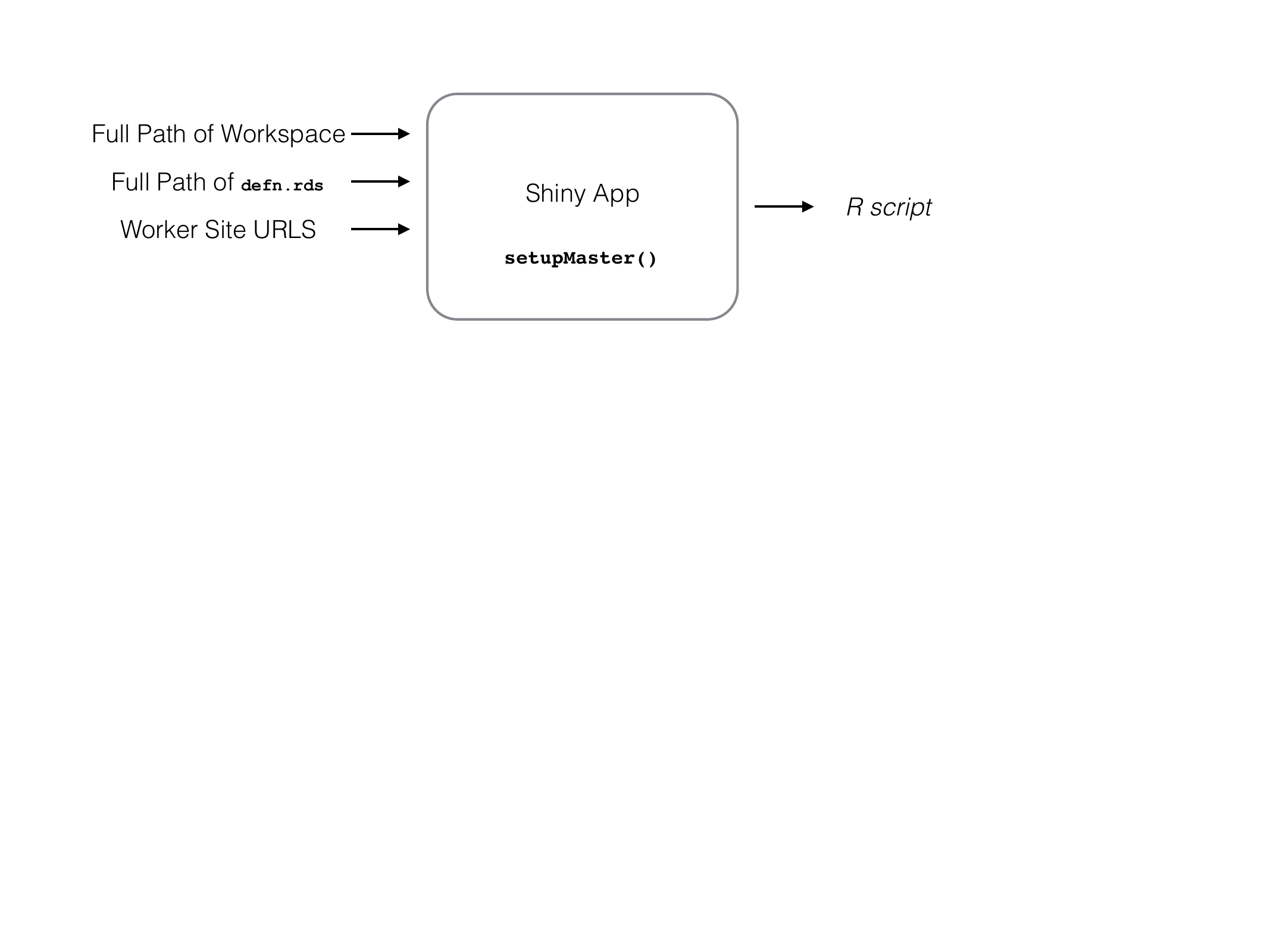}}
  \caption{Inputs, outputs, and associated \proglang{R} functions in package
    \pkg{distcomp} for each task.}
  \label{fig:steps}
\end{figure}

\section{Management of computations for a collaboration}
\label{sec:management}

We begin with an overview of the process and our software, with
details in the implementation section following. As illustrated in
Figure~\ref{fig:steps}, the setup of the software to manage
computations for a collaboration begins with a few steps that need
only be completed once at at a site, and that will create an
environment that will support many different collaborative
efforts. The local systems administrator must provide a computation
server, install the application \pkg{distcomp}, and make the server
URL available to the site investigators. The administrator would also
set up the required permissions.

The other steps are taken by the site investigator for a given
collaborative project (a ``computation''). We suppose first that the
investigator is the originator of a new computation and therefore, has
access to a dataset that will be used as that site's contribution to
the shared effort. The investigator runs a function from the
\pkg{distcomp} package on his or her local computer, which brings
up a browser window with instructions for entering the location of the
dataset, the names of its variables, and certain information about the
format and values of the fields. The formula, in \proglang{R} format is
entered. Completing this task and surviving various checks on data
integrity generates a unique computation-specific identifier and a
collection of metadata that define the computation and dataset.

Then the investigator sets up the worker process by running another
function in the \pkg{distcomp} package. The result is that a copy
of the local dataset and the metadata generated in the previous step
is placed in the computation server under the unique computation
identifier. At this point the originating investigator is ready to
recruit other collaborators. They will have to take the step of
creating a computation server, as described above, unless it has
already been done at that site. Then, they will run the same setup
function locally, and enter the computation definition (metadata) sent
by the originating investigator. The function will return a URL
(pointing to the computation server) that is to be transmitted back to
the originating investigator.

Once the originating investigator has the URLs from the sites in
hand, he or she runs the function for setting up a master site, and
then finally another function that runs the computation. The process
of setting up worker and master sites can be repeated for any number of
different computations and datasets.

We imagine that a group of collaborators interested in a particular
topic (say, prediction of outcome of therapy for a subtype of
melanoma) would run (and subsequentially update) several such
computations, perhaps variations of models.  The distributed
computations of interest to that group investigators might be listed
and managed on a website where a registry of active and proposed
projects is maintained.  Depending on the scope of the collaboration
there may be several such websites. One may imagine a website
providing the following capabilities for a collaboration:

\begin{description}
\item[Registration] Registering users, providing credentials,
  bonafides etc.
\item[Available Methods] A mechanism for adding to a growing list of
  statistical computations that may be implemented in the
  \pkg{distcomp} projects. The latest version of the package will
  automatically provide an up-to-date list.
\item[Project Linkages] Associating users with projects so that they
  may participate in the collaboration. In order to prevent malicious
  use and avoid inappropriate computational burdens on already
  established projects there are some checks and balances, such as
  requiring a new user to bring new data to the collaboration.
\item[Project Examples/Templates] Providing some testable examples for
  potential collaborators so that there is a complete understanding of
  the requirements of each site. Some canonical templates will be
  provided. Users may run them on local sites to completely examine
  the entire process.
\item[Project Publishing] Creation of projects and publishing them so
  that they move into place on a publicly visible website. A group of
  users or a single user will create a new collaborative project and
  lead it.
\item[Tools] Downloadable tools, virtual machines and detailed tool
  documentation for enabling dryrun experiments locally.
\item[Documentation] Wikis for facilitating communication within the
  collaboration.
\end{description}

There are many open source tools that can be used for some of the
tasks described above. We focus our attention on those tasks for which
new tools required development. These are the tools that allow rapid
setup of a computation, at both master and worker sites, as described
above, with minimal ongoing burden once the initial steps are
taken. The intent is to lower the bar to collaboration to the point
where it is as easy to set up a distributed collaboration as it would
be to even begin the discussion of data aggregation.

\section{Illustration}
\label{sec:illustration}

To get a sense of what is being transmitted between a master and
worker sites, consider a setting where several distributed sites have
data on patient age and one wishes to compute the mean age of all
patients. Assuming each site is willing to share the sum $\bar{X}_i$
of the patient ages at their site along with the number of patients
$n_i$, a master process could calculate the overall mean by requesting
the pair $(\bar{X}_i, n_i)$ from each site and computing
$\bar{X} = \sum_i n_i\bar{X}_i/\sum_in_i.$ Only the summaries
$(\bar{X}_i, n_i)$ are transmitted by the sites to the master but not
actual patient level data. This sort of computation is also
reminiscent of meta-analysis, where data from several studies gets
pooled to obtain meta-analytic estimates.

The situation is only slightly more involved in model fitting.

We focus on survival data, where the number of covariates might be
gathered on subjects that are followed over time. The random quantity
of interest is the time to ``event'' $T_i$ for each subject $i$ which
may be observed in some patients (in which case the observation is
uncensored) or not (in which case it is censored). Thus, each subject
$i$ yields data $(T_i, \delta_i)$ where $\delta_i=1$ if event, 0
otherwise, independent of $T_i$ in addition to covariates that may
vary over time.

The Cox proportional hazards model assumes a hazard function of the
form
\begin{equation}
  \label{eq:prop-haz}
  {\boldLambda}(t) = \lambda_0(t)\exp({\boldX}{\boldBeta}),
\end{equation}
where $\boldLambda$ is an $n\times 1$ vector, $n$ being the number of
subjects, $\lambda_0$, an unspecified baseline hazard, $\boldX$ the
$n\times p$ vector of covariates and $\boldBeta$ a $p\times 1$ vector
of coefficients.

Model fitting and inference is accomplished by maximizing a partial
likelihood function. The explicit form of this likelihood can be
written out using a counting process formulation following \citet{survival-book}.

For each subject $i$, let $N_i(t)$ be the number of observed events in
$[0, t]$ and $Y_i(t)$, the indicator of whether the subject is in the
``risk set'', that is, the subject may potentially contribute an event
at time $t$. Then the log of the partial likelihood function
introduced by \citet{cox1972} has the form:

\begin{equation}
  \label{eq:loglik}
  l(\boldBeta|\boldX) =  \sum_{i=1}^n \int_0^\infty
  \biggl[Y_i(t){\boldX}_i(t){\boldBeta} -
  \log\biggl(\sum_jY_j(t)r_j({\boldBeta}, t)\biggr)\biggr]dN_i(t)
\end{equation}
where $i$ indexes the subject, $r_i({\boldBeta}, t) =
\exp[{\boldX}_i\boldBeta]$ is the risk score, ${\boldX}_i(t)$ is the
covariate row vector. The score vector is given by:
\begin{equation}
  \label{eq:score}
  S(\boldBeta|\boldX) = \sum_{i=1}^n\int_0^\infty \boldZ_i(\boldBeta, s) dN_i(s)
\end{equation}
where
\begin{equation}
  \label{eq:zBeta}
  \boldZ_i(\boldBeta, s) = \boldX_i(s) - \overline{x}(\boldBeta, s)
\end{equation}
and
\begin{equation}
  \label{eq:xbar}
  \overline{x}(\boldBeta, s) = \frac{\sum Y_i(s)r_i(\boldBeta,s)\boldX_i(s)}
  {\sum Y_i(s)r_i(\boldBeta,s)}.
\end{equation}
The negative of the hessian is the Fisher information given by
\begin{equation}
  \label{eq:information}
  {\pmb I}(\boldBeta|\boldX) = \sum_{i=1}^n\int_0^\infty V(\boldBeta,s)dN_i(s)
\end{equation}
where
\begin{equation}
  \label{eq:1}
  V(\boldBeta, s) = \frac{\sum Y_i(s)r_i(\boldBeta,s)\bigl[\boldZ_i^{\top}(\boldBeta,
    s)\boldZ_i(\boldBeta, s)\bigr]}
  {\sum Y_i(s)r_i(\boldBeta,s)}.
\end{equation}

\begin{figure}
  \centering
  \includegraphics[width=0.63\textwidth]{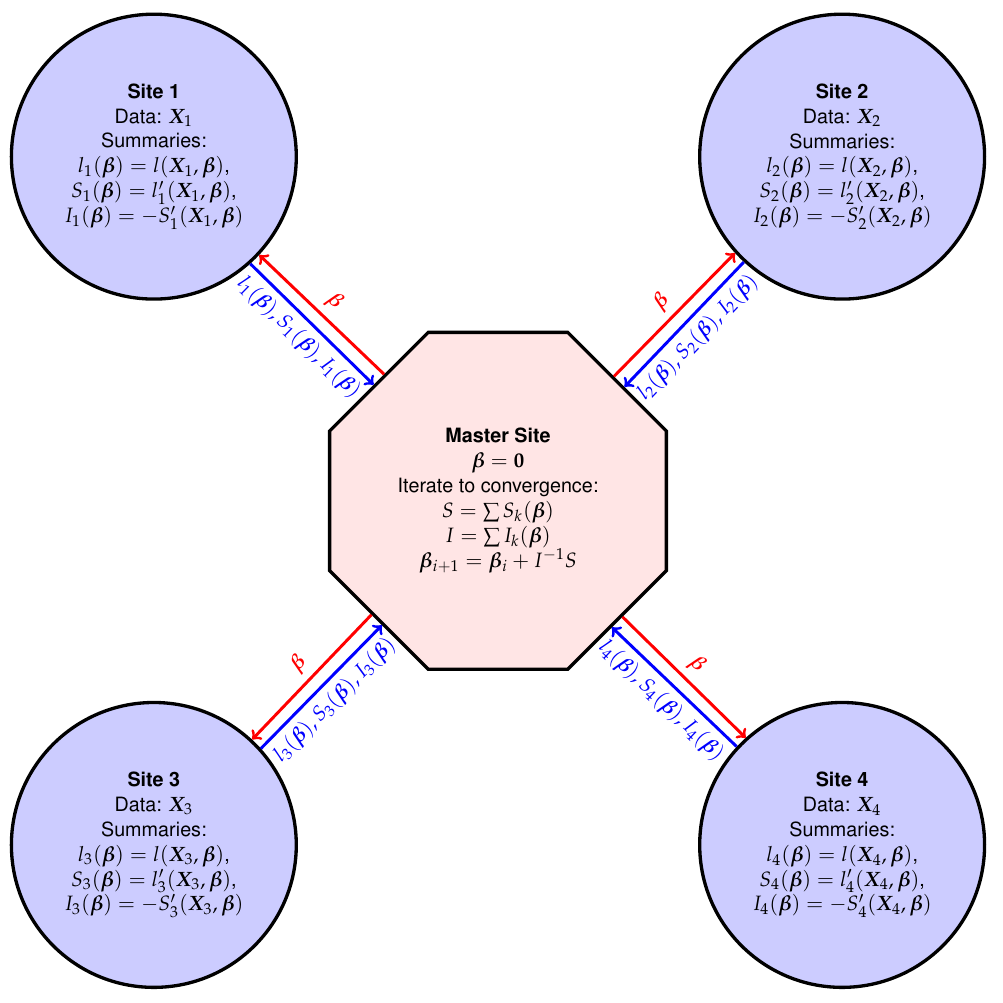}
  \caption{Summary quantities transmitted between master and worker
    sites in a stratified Cox model fit involving $K=4$ sites. The red
    arrows show what the master sends to each site and the blue arrows
    indicate what the sites return back. The summaries $l$, $S$ and
    $I$ are per equations~\ref{eq:liksum}, \ref{eq:scoresum},
    \ref{eq:fishersum} respectively.}
  \label{fig:hub}
\end{figure}

The solution $\hat{\boldBeta}$ to the score Equation~\ref{eq:score}
and the inverse of the information matrix $I^{-1}(\hat{\boldBeta})$
are used as the estimate and the variance respectively along with the
fact that the estimate is asymptotically normally distributed. One can
use a Newton-Raphson method to solve Equation~\ref{eq:score} with an
initial value for $\boldBeta$, often just $\pmb 0$. The method
is quite robust and in fact implemented in the popular \proglang{R}
\code{survival} package~\citep{survival-package}.

We are interested in \emph{stratified} Cox models where the data is
divided into several strata, each with its own baseline hazard, yet
use a common set of coefficients $\boldBeta$.  Specifically, if there
are $K$ strata and one views the entire $n\times p$ matrix $\boldX$ as
a stacking of $k$ sub-matrices $\boldX^{[k]}$ each of dimension
$n_k\times p$ with $n=\sum_{k=1}^Kn_k$, then subject $i$ in stratum
$k$ incurs a hazard $\lambda_k(t)\exp(\boldX_i^{[k]}\boldBeta).$ Such
models are particularly applicable in multicenter studies and trials
where the one needs to account for patient mix at each institution,
for example. In this case, the overall log-likelihood is a sum over
the strata
\begin{equation}
  \label{eq:liksum}
  l(\boldBeta|\boldX) = \sum_{k=1}^K l(\boldBeta, \boldX^{[k]}).
\end{equation}
where each component of the sum is in fact given by
Equation~\ref{eq:loglik}.  It follows that the score function and the
Fisher information also partition into sums:
\begin{equation}
  \label{eq:scoresum}
  S(\boldBeta|\boldX) = \sum_{k=1}^KS(\boldBeta, \boldX^{[k]})
\end{equation}
and
\begin{equation}
  \label{eq:fishersum}
  I(\boldBeta|\boldX) = \sum_{k=1}^KI(\boldBeta, \boldX^{[k]})
\end{equation}
where the components in each sum above are given by
Equations~\ref{eq:score} and \ref{eq:information}.

Note that in each stratum $k$, the computation of the likelihood,
score and information for the site-stratified Cox proportional hazards
model uses only stratum data $\boldX^{[k]}$. This feature of the
site-stratified model enables distributed computation, since each site
in a distributed optimization of the partial likelihood only need
provide values of $l(\boldBeta, \boldX^{[k]})$,
$U(\boldBeta|\boldX^{[k]})$, and $I(\boldBeta|\boldX^{[k]})$ for a
master process to estimate $\boldBeta$ and its variance. This is
illustrated in Figure~\ref{fig:hub} where four sites participate in a
stratified Cox fit.

\section{Implementation}
\label{sec:implementation}

Our implementation is in the form of the \pkg{distcomp} \proglang{R}
package which utilizes other \proglang{R} packages notably
\pkg{opencpu}~\citep{ooms2014} and \pkg{shiny}~\citep{shiny}, to
provide the infrastructure support. (The package is on the
Comprehensive \proglang{R} Archive Network and may therefore be installed like
any other.) We use a master-worker model, where the master is the one
in charge of the overall computation (the main iteration in an
optimization for example) and the workers merely compute functions on
local datasets and return the function result. Both the master and
worker sites are expected to have our package \pkg{distcomp}
installed.  In the course of a single fit, the master process will
make an unspecified number of computation requests to the worker sites
over secure \code{http} protocol.

In what follows, we assume an \pkg{opencpu} server for distributed
computations.

We describe our implementation first in the context of a distributed
stratified Cox Model fit discussed in Section~\ref{sec:illustration}
using the UIS dataset from \citet{asa} on time until return to drug
use for patients enrolled in two different residential treatment
programs.  Assuming all data in one place, one would fit a stratified
cox proportional hazard model using \code{site} (0 or 1) as a
stratification variable as follows.

\begin{CodeOutput}
R> uis <- readRDS("uis.RDS")
R> coxOrig <- coxph(formula = Surv(time, censor) ~ age +
+    becktota + ndrugfp1 + ndrugfp2 + ivhx3 + race +
+    treat + strata(site), data = uis)
R> summary(coxOrig)
Call:
coxph(formula = Surv(time, censor) ~ age + becktota + ndrugfp1 +
ndrugfp2 + ivhx3 + race + treat + strata(site), data = uis)

n= 575, number of events= 464
(53 observations deleted due to missingness)

               coef exp(coef)  se(coef)      z Pr(>|z|)
age       -0.028076  0.972315  0.008131 -3.453 0.000554 ***
becktota   0.009146  1.009187  0.004991  1.832 0.066914 .
ndrugfp1  -0.521973  0.593349  0.124424 -4.195 2.73e-05 ***
ndrugfp2  -0.194178  0.823512  0.048252 -4.024 5.72e-05 ***
ivhx3TRUE  0.263634  1.301652  0.108243  2.436 0.014868 *
race      -0.240021  0.786611  0.115632 -2.076 0.037920 *
treat     -0.212616  0.808466  0.093747 -2.268 0.023331 *
---
Signif. codes:  0 '***' 0.001 '**' 0.01 '*' 0.05 '.' 0.1 ' ' 1
...
\end{CodeOutput}

We now aim to reproduce the same results using a distributed
computation using \pkg{opencpu}. In order for the reader to reproduce
our example, we use the same \pkg{opencpu} server on a single machine
to simulate different sites; the package code automatically detects
such use to keep the site-specific data separate.

Some setup is involved before proceeding: an empty directory has to be
set aside for the workspace for the \pkg{opencpu} server. The details
and the structure of the workspace are further described in
Appendix~\ref{sec:ui}.

\begin{description}
\item[Define the computation] We define a computation definition
  object (\emph{compdef} for brevity) that will encode the
  characteristics of the computation. It contains an identifier along
  with a formula (therefore, the variables) that will be used in the
  model. Note that no data is stored in the definition, only the
  characteristics of the computation. This \emph{compdef} serves as an
  unambiguous definition of a computation task, which, in this case,
  happens to be a particular instance of a stratified Cox model using
  the variables specified in the formula.

\begin{CodeInput}
R> coxDef <- data.frame(compType = names(availableComputations())[1],
+    formula = paste("Surv(time, censor) ~ age + becktota + ",
+                   "ndrugfp1 + ndrugfp2 + ivhx3 + race + treat"),
+    id = "UIS", stringsAsFactors = FALSE)
\end{CodeInput}

\item[Set up worker processes for the computation] We will split the
  data by site and set up worker processes using \pkg{opencpu}. Then
  we upload the \emph{compdef} along with the site-appropriate data to
  each worker.

\begin{CodeInput}
R> library("opencpu")
R> siteData <- with(uis, split(x = uis, f = site))
R> nSites <- length(siteData)
R> sites <- lapply(seq.int(nSites),
+    function(x) list(name = paste0("site", x),
+    url = opencpu$url()))
R> ok <- Map(uploadNewComputation, sites,
+    lapply(seq.int(nSites), function(i) coxDef), siteData)
R> stopifnot(all(as.logical(ok)))
\end{CodeInput}

At this point, the two sites in this example each have access to their
private data and the computation definition. Since they each also have
the \pkg{distcomp} installed (by default in this case), they also have
the code required to engage in the computation.  If either the data or
the variables were incompatible, an error would have been raised.

\item[Build a master process for the computation] Once the sites are
  set up, the master object for the computation can be constructed
  using the \emph{compdef} identifier and the formula. Each site
  worker is represented by the \pkg{opencpu} web address and these are
  added to the master so that the master may use them in performing
  the computation.

\begin{CodeInput}
R> master <- CoxMaster$new(defnId = coxDef$id,
+    formula = coxDef$formula)
R> for (site in sites) {
+    master$addSite(name = site$name, url = site$url)
+  }
\end{CodeInput}

\item[Fit the model] At this point, the master is ready to perform the
  computation, which in this case is maximizing the Cox partial
  likelihood. Calling the \code{run} method of the master performs
  this optimization and the resulting summary can be printed out.

  \begin{CodeOutput}
R> result <- master$run()
R> print(master$summary(), digits = 5)
        coef exp(coef)  se(coef)       z          p
1 -0.0280495   0.97234 0.0081301 -3.4501 5.6041e-04
2  0.0091441   1.00919 0.0049918  1.8318 6.6979e-02
3 -0.5219296   0.59337 0.1244240 -4.1948 2.7315e-05
4 -0.1941709   0.82352 0.0482507 -4.0242 5.7168e-05
5  0.2636376   1.30166 0.1082448  2.4356 1.4868e-02
6 -0.2400609   0.78658 0.1156319 -2.0761 3.7887e-02
7 -0.2125720   0.80850 0.0937466 -2.2675 2.3359e-02
\end{CodeOutput}

  As can be seen, the results are similar to the original model fit.

\end{description}

In the above example, the master performs an optimization of a
multivariate likelihood function that is the sum of the likelihoods at
each site as shown in Equation~\ref{eq:liksum}. Using a starting value
for the parameter $\boldBeta = 0$, it repeatedly sends updated
$\boldBeta$ values until a convergence criterion is reached.

We have successfully tested other examples in a real three-site
configuration, involving two US and one UK site. Tests involving
real locally-originated data are underway.

\section{A rank-$k$ approximation example}
\label{sec:rank-k-example}

We consider the problem of approximating a matrix by another low-rank
matrix. Assuming that the original matrix $\boldX$ is row-partioned
into sub-matrices $\boldX_j$ at $j=1\ldots,K$ sites, there is a
well-known iterative singular value decomposition algorithm (see
Appendix~\ref{sec:rank-k}) to obtain a low-rank approximation using
the singular vectors, which is implemented in \pkg{distcomp}. The
example below assumes three sites.

\begin{CodeOutput}
R> print(availableComputations())
$StratifiedCoxModel
$StratifiedCoxModel$desc
[1] "Stratified Cox Model"
...
$RankKSVD
$RankKSVD$desc
[1] "Rank K SVD"
...
\end{CodeOutput}

Using this information, we construct the \emph{compdef} for this
computation where we will compute the first two singular values for a
five-column matrix and use a generic identifier \code{SVD}.
\begin{CodeInput}
R> svdDef <- data.frame(compType = names(availableComputations())[2],
+    rank = 2L,
+    ncol = 5L,
+    id = "SVD",
+    stringsAsFactors = FALSE)
\end{CodeInput}
We now generate random data for three sites, start the \pkg{opencpu}
server, and set the URLs for the sites to be the (same) local
\pkg{opencpu} URL.
\begin{CodeInput}
R> set.seed(12345)
R> nSites <- 3
R> siteData <- lapply(seq.int(nSites),
+    function(i) matrix(rnorm(100), nrow = 20))
R> library("opencpu")
R> sites <- lapply(seq.int(nSites),
+    function(x) list(name = paste0("site", x),
+    url = opencpu$url()))
\end{CodeInput}

Next, we upload the data to the three sites, ensuring that we use
different names for the data files for each site.
\begin{CodeInput}
R> ok <- Map(uploadNewComputation, sites,
+    lapply(seq.int(nSites), function(i) svdDef), siteData)
R> stopifnot(all(as.logical(ok)))
\end{CodeInput}
At this point, the sites are instantiated and ready to compute. So we
instantiate the master and add the three participating sites.

\begin{CodeInput}
R> master <- SVDMaster$new(defnId = svdDef$id, k = svdDef$rank)
R> for (site in sites) {
+    master$addSite(name = site$name, url = site$url)
+  }
\end{CodeInput}

All that remains is to call the \code{run} method of the master
object.
\begin{CodeOutput}
R> result <- master$run()
...
R> print(result)
$v
            [,1]        [,2]
[1,]  0.17947030  0.08275684
[2,]  0.78969198  0.34634459
[3,] -0.21294972  0.91875219
[4,] -0.54501407  0.16784298
[5,]  0.04229739 -0.03032954

$d
[1] 9.707451 8.200043
\end{CodeOutput}
This returns the approximate first two singular values and the associated
vectors (up to a sign change). All five singular vectors can be
obtained as shown below. We also compare the results to the SVD on the
aggregated matrix.
\begin{CodeOutput}
R> result <- master$run(k = 5)
...
R> x <- do.call(rbind, siteData)
R> print(result$d)
[1] 9.707451 8.200043 7.982650 7.257355 6.235351
R> print(svd(x)$d)
[1] 9.707537 8.199827 7.982888 7.257286 6.235182
R> print(result$v)
            [,1]        [,2]       [,3]        [,4]         [,5]
[1,]  0.17947030  0.08275684  0.0165604  0.98008722 -0.008933396
[2,]  0.78969198  0.34634459 -0.3437723 -0.16504730  0.333181988
[3,] -0.21294972  0.91875219  0.2496210 -0.04479619 -0.214978886
[4,] -0.54501407  0.16784298 -0.5334277  0.10025749  0.616612820
[5,]  0.04229739 -0.03032954  0.7312254 -0.01140918  0.680060781
R> print(svd(x)$v)
            [,1]        [,2]        [,3]        [,4]        [,5]
[1,] -0.17946375  0.08268613 -0.01644895 -0.98010572 -0.00883063
[2,] -0.78963831  0.34694371  0.34328503  0.16509457  0.33316749
[3,]  0.21305901  0.91839439 -0.25083926  0.04461477 -0.21505068
[4,]  0.54504905  0.16843629  0.53318714 -0.10009622  0.61663844
[5,] -0.04232602 -0.03120945 -0.73121540  0.01126215  0.68002329
\end{CodeOutput}

As we can see, the distributed~SVD is able to recover the same factors
(up to sign change) as the standard SVD algorithm.  There is a slight
loss of numerical precision in the alternating power algorithm versus
the standard LAPACK implementation. The JSON serialization format
employed here also loses precision. Future implementations will use a
more portable format such as Google protocol buffers, as we note in
Section~\ref{sec:sec-safe}.

\section{Privacy control and confidence}
\label{sec:sec-safe}
The motivating principles guiding the architecture of the distributed
computation method we have built include the preservation of data
privacy, retention of local control and building confidence in the
safety of the method by using an open platform. We list several
features that serve that principle. Those features were chosen based
on a particular threat model that deals with internal threats, since
an anonymous entity is not a participant in the collaborative model
fitting. We expand on these below.

\begin{description}
\item[Registering machines] Only known machines are permitted to
  request computations. In our model, the master and workers are
  trusted; this comes about as a result of the initial process in
  agreeing to collaborate on a computation. This trust also defines
  and limits the threat model for us, namely, that of a rogue
  investigator trying to peek at intermediate results and trying to
  break the system, i.e., an internal threat. The audit tools
  described below in analyzing logs will aid in limiting such a
  threat.
\item[Logging] All requests for computation are logged by the local
  computation server, providing accountability. Site specific
  dashboards can be developed (as \pkg{shiny} apps, for example) for
  analyzing these logs and flagging unusual requests.
\item[Single computation server] A site could use a single computation
  server for all distributed collaborations, or a small number,
  simplifying the oversight task. At any site, it is typically an IT
  team that really handles access to the public facing web
  resources. In such a case, there is usually a single web URL (or
  even server, for that matter) serves as a gateway for all
  \texttt{distcomp} computations for a worker site.  This
  \emph{gatekeeper} usually has firewall rules installed so that only
  authorized masters can make computation queries.  This simplicity,
  however, creates a loophole. Our model relies on associating a
  unique identifier with each \emph{compDef} and associated data at
  each site. Thus, if worker site A participates in two computations
  \code{id1} and \code{id2} with two different institutions B and C
  respectively, there is the possibility that B may be able to run a
  computation on data that was made available to site C and vice-versa
  if B knows \code{id2}. However, this is easily handled by means of
  an access rule that maps each master to a set of computations the
  master can access via a reverse proxy~\citep{wiki_reverseproxy}
  application either in the gatekeeper or in front of the
  \pkg{opencpu} server ensuring proper access.

\item[Contingent participation] A site agrees to or refuses a specific
  computation, which it can inspect, and can withdraw at any time from
  any computation, so it retains complete control over the future use
  of its data.
\item[Limited computation] The computation server can only respond to
  the specific computation request described in the \emph{compdef} and
  also only those exposed by \pkg{distcomp}. This can be implemented
  by accepting requests for web addresses only of a particular
  form---a standard technique---at the the gatekeeper application thus
  preventing other functions or scripts are from being executed.
\end{description}

\section{Some possible use cases}
\label{sec:use-cases}

We anticipate that the main initial use will be to make it possible
for investigators who are already known to each other to quickly pool
information by mounting "pop-up" collaborations. Once the system is
set up, understood and has passed local due diligence for privacy
protection, the additional cost and effort required to start a new
computation among such a working group should be negligible. As use
cases occur, a byproduct will be a growing set of mutually understood
data fields, making future computations easier to set up.

There are a few other use cases that we expect to see after some time.

\paragraph{Preparation for aggregation.} A group of sites who are
considering the possibility of aggregating data for some purpose might
start their collaboration with a few "small wins", to test their
ability to rationalize their data dictionaries and see if there is
sufficient value in aggregation.

\paragraph{Combining information when control over data use is important.}
Owners of data, such as pharmaceutical companies, who are not willing
to transfer them may be willing to participate in collaborations that
make use of the distributed technologies. In a single-worker
configuration, information from a single source could be used to fit a
model proposed by an outside party (acting as the master). This
provides a way for organizations to allow others to access their data
(for the purpose of model fitting) in a controlled manner.

\paragraph{Virtual aggregation.} In a group of sites that often collaborate,
a library of "translated terms" and the associated data fields could
be developed, creating a virtual data aggregation, whose information
would be accessed by distributed model building. If sufficient trust
has been created, it is possible to make the setup of a new
computation nearly transparent to the investigator.

\section{Discussion}
\label{sec:discussion}

The data aggregator must work in a highly restrictive regulatory
regime that controls and limits the export of (identified) Protected
Health Information (PHI) beyond the originating institution. Owners of
data can be reluctant to cede control over the use of their data, even
when privacy issues can be resolved. It seems realistic to predict
that regulatory concerns, privacy issues, and reluctance to let raw
data leave the local source will continue to make it difficult to
construct central repositories of data on biomarkers, treatments, and
clinical outcomes, despite ongoing efforts by the NIH to support (and
even mandate) such repositories. The barriers to access for such data
do not seem to be falling. This is of special concern for the
aggregators of data (such as gene sequences) that are increasingly
likely to emerge as part of the diagnostic and treatment process, and
therefore become part of a system of medical records. Such data would
be regulated as PHI (under HIPAA) in the future if a consensus
develops that they cannot be irreversibly de-identified.

The lack of uniform standards for data management, database
ontologies, and the specific content and scope of databases makes
global aggregation of complex datsets laborious, expensive, and
time-consuming. Reluctance to adopt common standards makes central
repositories difficult to achieve, and the effort is seldom funded by
outside sources. Investigators might be willing to pay the modest
price for standardizing the part of the data that is needed for a
particular project, and then to continue to increase common
ontological themes as they are rewarded with analyses.

There are limitations to our current implementation. We have not yet
built methods beyond the two described above. We are working on
others, including the \emph{lasso}. Each new modeling technique
requires the creation of a definition that contains the relevant
parameters, associated code to compute the summary statistics at the
worker sites, a matching master that makes use of the facilities and
registering it as an available computation in \pkg{distcomp}.  This
may seem like a vast task given the variety of techniques and
models. However, these methods fall broadly into a few categories: a
large and useful group that aggregate likelihoods and other
statistics, a large group that optimize a criterion, even in a
distributed manner like ADMM~\citep{admm}, etc. They all break up in a
manner amenable to our approach. The well-organized \proglang{R}
software is eminently suited to such implementations; often, new
methods can be implemented by mere re-engineering of already existing
(open) algorithms in \proglang{R} packages to enable distributed computations. We
expect that the repertoire of models will expand as others add their
own features and make them available. There is not yet an easy-to-use
procedure for handling factor variables whose support varies by
site. This and many other improvements in ease of use will require
further effort. It is often the case that most data at sites is in
databases such as Redcap or i2b2. Our examples have used pure
\proglang{R} structures for persisting the data. A simple modification
can enable the use of a system such as Redcap. In discussions with
personnel at some sites, this configuration seems most acceptable to
the CIOs, enabling periodic updates of the data.

We should mention, however, that one important advantage is offered by
our use of \pkg{opencpu} which comes in two flavors, a cloud
server---the one that real sites will use---and a local \proglang{R}
package---one that developers will use. This means that developers can
prototype and debug distributed computations with very little effort
on a laptop. However, a robust implementation of a distributed
computation must deal with many modes of failure. More work remains to
be done on this, particularly in implementing a robust messaging
system to deal with failures.

Any distributed computation must address the potential for race
conditions. We account for this in two ways in our package: (a) by
using instance objects so that even the same computation requests
initiated at two different time points (regardless of origin) are
guaranteed different instance objects and (b) using the master in a
sequential iteration so that a new request is not sent until either
the current request returns a result or times out. That said, there
may be other race conditions that can occur at the \pkg{opencpu} level
that may be discovered over time.

The current implementation makes uses of the JSON serialization
format, simply because it is the best supported format of
\pkg{opencpu}. It would be better to use a well-established portable
and stable serialization format, such as Google protocol
buffers~\citep{goog-pbuff, edd2014, rprotobuf2014}. It is
straightforward to adapt to this protocol once it is more widely
supported in \pkg{opencpu} for example.

These limitations of our first implementation notwithstanding, we hope
that lowering the barriers to shared statistical computation will
accelerate the pace of collaboration and increase the accessibility of
information that is now locked up.

\section{Acknowledgements}
\label{sec:acknowledgements}

This work by supported in part by grants the Cancer Center Support
grant 1 P30 CA124435 from the National Cancer Center, the Clinical and
Translational Science Award 1UL1 RR025744 for the Stanford Center for
Clinical and Translational Education and Research (Spectrum) from the
National Center for Research Resources, National Institutes of Health
and award LM07033 from the National Institutes of Health.

\bibliography{jss1545}

\begin{thebibliography}{22}
\newcommand{\enquote}[1]{``#1''}
\providecommand{\natexlab}[1]{#1}
\providecommand{\url}[1]{\texttt{#1}}
\providecommand{\urlprefix}{URL }
\expandafter\ifx\csname urlstyle\endcsname\relax
  \providecommand{\doi}[1]{doi:\discretionary{}{}{}#1}\else
  \providecommand{\doi}{doi:\discretionary{}{}{}\begingroup
  \urlstyle{rm}\Url}\fi
\providecommand{\eprint}[2][]{\url{#2}}

\bibitem[{Anderson \emph{et~al.}(1999)Anderson, Bai, Bischof, Blackford,
  Demmel, Dongarra, Du~Croz, Greenbaum, Hammerling, McKenney
  \emph{et~al.}}]{anderson1999lapack}
Anderson E, Bai Z, Bischof C, Blackford S, Demmel J, Dongarra J, Du~Croz J,
  Greenbaum A, Hammerling S, McKenney A, \emph{et~al.} (1999).
\newblock \emph{LAPACK Users' guide}, volume~9.
\newblock Siam.

\bibitem[{Boyd \emph{et~al.}(2011)Boyd, Parikh, Chu, Peleato, and
  Eckstein}]{admm}
Boyd S, Parikh N, Chu E, Peleato B, Eckstein J (2011).
\newblock \enquote{Distributed Optimization and Statistical Learning via the
  Alternating Direction Method of Multipliers.}
\newblock \emph{Found. Trends Mach. Learn.}, \textbf{3}(1), 1--122.
\newblock ISSN 1935-8237.
\newblock \doi{10.1561/2200000016}.
\newblock \urlprefix\url{http://dx.doi.org/10.1561/2200000016}.

\bibitem[{Chang \emph{et~al.}(2015)Chang, Cheng, Allaire, Xie, and
  McPherson}]{shiny}
Chang W, Cheng J, Allaire J, Xie Y, McPherson J (2015).
\newblock \emph{shiny: Web Application Framework for R}.
\newblock R package version 0.11.1,
  \urlprefix\url{http://CRAN.R-project.org/package=shiny}.

\bibitem[{Chen \emph{et~al.}(2013)Chen, Martin, Daimon, and
  Maudsley}]{chen2013effective}
Chen H, Martin B, Daimon CM, Maudsley S (2013).
\newblock \enquote{Effective use of latent semantic indexing and computational
  linguistics in biological and biomedical applications.}
\newblock \emph{Frontiers in physiology}, \textbf{4}.

\bibitem[{Cox(1972)}]{cox1972}
Cox DR (1972).
\newblock \enquote{{Regression Models and Life-Tables}.}
\newblock \emph{Journal of the Royal Statistical Society. Series B
  (Methodological)}, \textbf{34}(2), 187--220.
\newblock
  \urlprefix\url{http://links.jstor.org/sici?sici=0035-9246\%281972\%2934\%3A2\%3C187\%3ARMAL\%3E2.0.CO\%3B2-6}.

\bibitem[{Eddelbuettel \emph{et~al.}(2014)Eddelbuettel, Stokely, and
  Ooms}]{edd2014}
Eddelbuettel D, Stokely M, Ooms J (2014).
\newblock \enquote{RProtoBuf: Efficient Cross-Language Data Serialization in
  R.}
\newblock \emph{Arxiv}.
\newblock \doi{arXiv:1401.7372}.

\bibitem[{Francois \emph{et~al.}(2014)Francois, Eddelbuettel, Stokely, and
  Ooms}]{rprotobuf2014}
Francois R, Eddelbuettel D, Stokely M, Ooms J (2014).
\newblock \emph{RProtoBuf: R Interface to the Protocol Buffers API. R package
  version}.
\newblock \doi{arXiv:1406.4806}.

\bibitem[{{Google}(2012)}]{goog-pbuff}
{Google} (2012).
\newblock \emph{Protocol Buffers: Developer Guide}.
\newblock \urlprefix\url{URL
  http://code.google.com/apis/protocolbuffers/docs/overview.html}.

\bibitem[{Hosmer \emph{et~al.}(2008)Hosmer, Lemeshow, and May}]{asa}
Hosmer D, Lemeshow S, May S (2008).
\newblock \emph{Applied Survival Analysis}.
\newblock John Wiley \& Sons.
\newblock ISBN-13: 0471754994,
  \urlprefix\url{http://www.wiley.com/WileyCDA/WileyTitle/productCd-0471754994.html}.

\bibitem[{Jiang \emph{et~al.}(2013)Jiang, Li, Wang, Wu, Xue, Ohno-Machado, and
  Jiang}]{jiang_webglore:_2013}
Jiang W, Li P, Wang S, Wu Y, Xue M, Ohno-Machado L, Jiang X (2013).
\newblock \enquote{{WebGLORE:} a web service for Grid {LOgistic} {REgression}.}
\newblock \emph{Bioinformatics (Oxford, England)}, \textbf{29}(24), 3238--3240.
\newblock ISSN 1367-4811.
\newblock \doi{10.1093/bioinformatics/btt559}.
\newblock {PMID:} 24072732 {PMCID:} {PMC3842761}.

\bibitem[{Leek and Storey(2007)}]{leek2007capturing}
Leek JT, Storey JD (2007).
\newblock \enquote{Capturing heterogeneity in gene expression studies by
  surrogate variable analysis.}
\newblock \emph{PLoS genetics}, \textbf{3}(9), e161.

\bibitem[{Lu \emph{et~al.}(2015)Lu, Wang, Ji, Wu, Xiong, Jiang, and
  Ohno-Machado}]{Luocv083}
Lu CL, Wang S, Ji Z, Wu Y, Xiong L, Jiang X, Ohno-Machado L (2015).
\newblock \enquote{WebDISCO: A web service for distributed cox model learning
  without patient-level data sharing.}
\newblock \emph{Journal of the American Medical Informatics Association}.
\newblock ISSN 1067-5027.
\newblock \doi{10.1093/jamia/ocv083}.

\bibitem[{Mazumder \emph{et~al.}(2010)Mazumder, Hastie, and
  Tibshirani}]{mazumder2010spectral}
Mazumder R, Hastie T, Tibshirani R (2010).
\newblock \enquote{Spectral regularization algorithms for learning large
  incomplete matrices.}
\newblock \emph{The Journal of Machine Learning Research}, \textbf{11},
  2287--2322.

\bibitem[{Murtagh \emph{et~al.}(2012)Murtagh, Demir, Jenkings, Wallace,
  Murtagh, Boniol, Bota, Laflamme, Boffetta, Ferretti, and
  Burton}]{murtagh_securing_2012}
Murtagh MJ, Demir I, Jenkings KN, Wallace SE, Murtagh B, Boniol M, Bota M,
  Laflamme P, Boffetta P, Ferretti V, Burton PR (2012).
\newblock \enquote{Securing the data economy: translating privacy and enacting
  security in the development of {DataSHIELD}.}
\newblock \emph{Public health genomics}, \textbf{15}(5), 243--253.
\newblock ISSN 1662-8063.
\newblock \doi{10.1159/000336673}.
\newblock {PMID:} 22722688.

\bibitem[{NIH(2015)}]{nih_collab}
NIH (2015).
\newblock \enquote{NIH Collaboratory: Rethinking Clinical Trials.}
\newblock Accessible at
  \url{https://www.nihcollaboratory.org/Pages/distributed-research-network.aspx}.

\bibitem[{Novembre \emph{et~al.}(2008)Novembre, Johnson, Bryc, Kutalik, Boyko,
  Auton, Indap, King, Bergmann, Nelson \emph{et~al.}}]{novembre2008genes}
Novembre J, Johnson T, Bryc K, Kutalik Z, Boyko AR, Auton A, Indap A, King KS,
  Bergmann S, Nelson MR, \emph{et~al.} (2008).
\newblock \enquote{Genes mirror geography within Europe.}
\newblock \emph{Nature}, \textbf{456}(7218), 98--101.

\bibitem[{Ooms(2014)}]{ooms2014}
Ooms J (2014).
\newblock \enquote{The OpenCPU System: Towards a Universal Interface for
  Scientific Computing through Separation of Concerns.}
\newblock \emph{Arxiv}.
\newblock \doi{arXiv:1406.4806}.
\newblock \urlprefix\url{http://www.opencpu.org}.

\bibitem[{{R Core Team}(2014)}]{r-core}
{R Core Team} (2014).
\newblock \emph{R: A Language and Environment for Statistical Computing}.
\newblock R Foundation for Statistical Computing, Vienna, Austria.
\newblock \urlprefix\url{http://www.R-project.org/}.

\bibitem[{{Terry M. Therneau} and {Patricia M. Grambsch}(2000)}]{survival-book}
{Terry M Therneau}, {Patricia M Grambsch} (2000).
\newblock \emph{Modeling Survival Data: Extending the {C}ox Model}.
\newblock Springer-Verlag, New York.
\newblock ISBN 0-387-98784-3.

\bibitem[{Therneau(2014)}]{survival-package}
Therneau TM (2014).
\newblock \emph{A Package for Survival Analysis in S}.
\newblock R package version 2.37-7,
  \urlprefix\url{http://CRAN.R-project.org/package=survival}.

\bibitem[{Wikipedia(2015)}]{wiki_reverseproxy}
Wikipedia (2015).
\newblock \enquote{Reverse Proxy.}
\newblock Accessible at \url{https://en.wikipedia.org/wiki/Reverse_proxy}.

\bibitem[{Wolfson \emph{et~al.}(2010)Wolfson, Wallace, Masca, Rowe, Sheehan,
  Ferretti, LaFlamme, Tobin, Macleod, Little
  \emph{et~al.}}]{wolfson2010datashield}
Wolfson M, Wallace SE, Masca N, Rowe G, Sheehan NA, Ferretti V, LaFlamme P,
  Tobin MD, Macleod J, Little J, \emph{et~al.} (2010).
\newblock \enquote{DataSHIELD: resolving a conflict in contemporary
  bioscience-performing a pooled analysis of individual-level data without
  sharing the data.}
\newblock \emph{International journal of epidemiology}, p. dyq111.

\end{thebibliography}

\appendix

\section{Distributed rank-$k$ singular value decompostion}
\label{sec:rank-k}

\begin{algorithm}[t]
  \begin{mdframed}
  \KwData{$\boldX \in \mR^{n\times p}$}
  \KwResult{$u \in \mR^{n}, \; v \in \mR^{p},$ and $d > 0$}
  $u \gets (\frac{1}{\sqrt{n}}, \frac{1}{\sqrt{n}}, \dots, \frac{1}{\sqrt{n}})$\;
  \Repeat{ convergence}{
    $v \gets \boldX^{\top}u$\;
    $v \gets v/\|v\|$\;
    $u \gets \boldX v$\;
    $d \gets \|u\|$\;
    $u \gets u/\|u\|$\;}
  \end{mdframed}
  \caption{Alternating algorithm for rank-1 SVD.}
  \label{fig:altsvd1}
\end{algorithm}

The singular value decomposition (SVD) is a method for summarizing a
dataset.  The SVD of $\boldX$, the $n\times p$ matrix of covariates,
is formed by the matrices $\pmb U$, $\pmb V$, $\pmb D$ such that
\begin{equation}
  \label{eq:svd}
  \pmb X = \pmb U \pmb D \pmb V^{\top},\; \pmb U^{\top} \pmb U = I,\; \pmb V^{\top} \pmb
  V = I,\; \textrm{and } \pmb D \textrm{ is diagonal with entries}.
\end{equation}

\begin{algorithm}
  \begin{mdframed}
  \KwData{each worker has private data $\boldX_i \in \mR^{n_i\times p}$}
  \KwResult{$V \in \mR^{p\times k},$ and $d_1 \ge \dots d_k \ge 0$}
  $V \gets 0, \; d \gets 0$
  \ForEach{worker site $j$}{
    $U^{[j]} = 0$\;
    transmit $n_j$ to master\;
  }
  \For{$i \gets 1$ \KwTo $k$}{
    \lForEach{worker site $j$}{$u^{[j]} \gets (1, 1, \dots, 1)$ of length $n_j$}
    $\|u\| \gets \sqrt{\sum_j n_j}$\;
    transmit $\|u\|, V,$ and $D$ to workers\;
    \Repeat{ convergence}{
      \ForEach{worker site $j$}{
        $u^{[j]} \gets u^{[j]}/\|u\|$\;
        calculate $v^{[j]} \gets (\boldX^{[j]} - U^{[j]}DV^{\top})^{\top}u^{[j]}$\;
        transmit $v^{[j]}$ to master\;
      }
      $v \gets \sum_j v^{[j]}$\;
      $v \gets v/\|v\|$\;
      transmit $v$ to workers\;
      \ForEach{worker site $j$}{
        calculate $u^{[j]} \gets \boldX^{[j]} v$\;
        transmit $\|u^{[j]}\|$ to master\;
      }
      $\|u\| \gets \sum_j \|u^{[j]}\|$\;
      transmit $\|u\|$ to workers\;
      $d_i \gets \|u\|$\;
    }
    $V \gets \texttt{cbind}(V, v)$\;
    \lForEach{worker site $j$}{$U^{[j]} \gets \texttt{cbind}(U^{[j]}, u^{[j]})$}
  }
  \end{mdframed}
  \caption{Privacy preserving algorithm for rank-$k$ SVD.}
  \label{fig:rankksvd}
\end{algorithm}

Additionally, we assume that the values of $\pmb D$ is nonnegative and
sorted, namely that $\pmb D = $\texttt{ diag($d_1,d_2, \dots,
  d_{\min(n,p)}$)} with $d_1 \ge d_2, \ge \dots \ge 0$.  With these
assumptions, the SVD of $\boldX$ is unique up to the signs of the
columns of $\pmb U$ and $\pmb V$.

The SVD is a sort of factor model which decomposes the variance of
$\boldX$ into what are called principal components.  $v_1$, the first
column of $\pmb V$, is called the first principal component of $X$ and
is the maximizer of $\mathrm{var}(\boldX v)$.  The values of $u_1$ can
be viewed as loading that tell you how much of the factor $v_1$ is
present in each observation.  $d_1^2/\sum_jd_j^2$ is the proportion of
the variance of $\boldX$ that can be explained by $v_1$, which means
that the SVD also sorts the components in order of their contribution
to variance.  When we take a rank-$k$ SVD we only keep the $k$ most
important factors, equivalent to setting $d_{k+1} = d_{k+2} = \dots =
d_{\min(n,p)} = 0$.

The SVD has many uses in medical research.  Often times, the results
are themselves of interest for examining a factor model.  This is the
case for methods like Latent Semantic Indexing
\citep{chen2013effective}.  In a recent example,
\citet{novembre2008genes} reconstruct a map of europe from the first
two principal components of SNP data.  Additionally, SVDs are often
used as a preprocessing step before some further data analysis.  There
are many examples of this, including Principal Component Regression,
SVD based imputation methods, and Surrogate Variable Analysis -- a
technique based on trying to remove unwanted variation from high
dimensional data \citep{mazumder2010spectral, leek2007capturing}.

Calculating an SVD can be a challenging problem from a numerical
stability standpoint.  For this reason, it is relatively rare for
people to implement their own version of an SVD; \proglang{R} and
\proglang{MATLAB} both use the LAPACK implementation
of \citet{anderson1999lapack}.  Most of these methods involve
transforming the $\boldX$ matrix in a way that cannot be done while
maintaining privacy.  That said, there is a relatively simple
iterative procedure for calculating the rank-1 SVD that can be
modified to be privacy preserving.  Algorithm~\ref{fig:altsvd1}
calculates $(u, v, d)$, the rank-1 SVD of $\boldX$.

Since singular vectors can be found successively by removing the
effect of the top singular vector and then finding the rank-1
approximation again, we can repeatedly apply the
algorithm~\ref{fig:altsvd1} in order to get a rank-$k$ SVD.  Combining
that idea with some modifications to the previous algorithm in order
to preserve privacy, we get algorithm~\ref{fig:rankksvd} for rank-$k$
privacy protecting SVD.

\begin{figure}[ht]
  \centering
  \includegraphics[width=.75\linewidth]{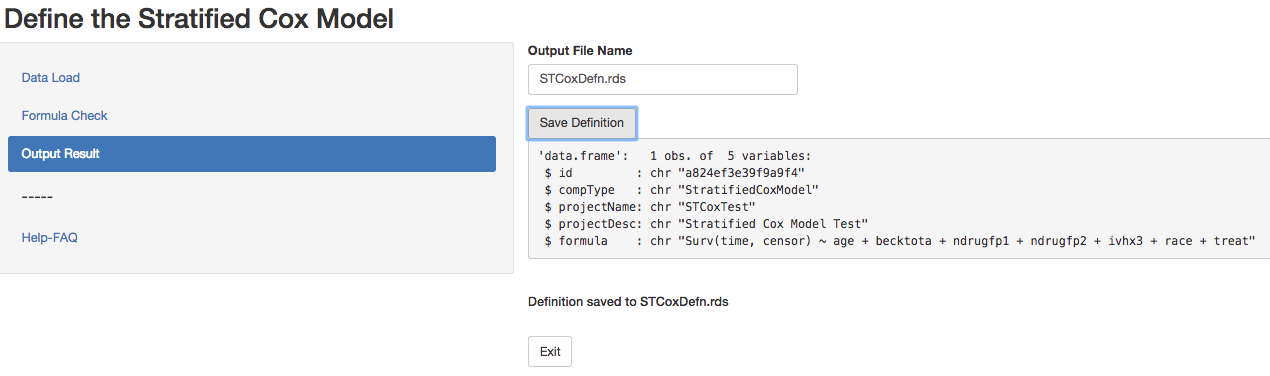}
  \caption{A part of defining a new computation.}
  \label{fig:shiny-define-comp}
\end{figure}

\begin{figure}[ht]
  \centering
  \tikzstyle{every node}=[thick,anchor=west, rounded corners,
  font={\ttfamily}, inner sep=2.5pt]
  \tikzstyle{selected}=[draw=blue,fill=blue!10]
  \tikzstyle{selected1}=[draw=blue,fill=blue!5]
  \tikzstyle{dots}=[draw=black,fill=none] \tikzstyle{root}=[selected,
  fill=blue!30]
\begin{tikzpicture}[%
  grow via three points={one child at (0.5,-0.7) and
  two children at (0.5,-0.7) and (0.5,-1.4)},
  edge from parent path={(\tikzparentnode.south) |- (\tikzchildnode.west)}]
\node [root] {workspace}
child { node [selected] {defn}
  child {
    node  [selected1] {defnId1}
    child { node {defn.rds} }
    child { node {data.rds} }
  }
  child [missing] {}
  child [missing] {}
  child {
    node [selected1] {defnId2}
    child { node {defn.rds} }
    child { node {data.rds} }
  }
  child [missing] {}
  child [missing] {}
  child {
    node {\vdots}
  }
}
child [missing] {}
child [missing] {}
child [missing] {}
child [missing] {}
child [missing] {}
child [missing] {}
child [missing] {}
child [missing] {}
child { node [selected] {instances}
  child { node {instance1.rds}}
  child { node {instance2.rds}}
  child {
    node  {\vdots}
  }
};
\end{tikzpicture}
  \caption{Structure of workspace area.}
  \label{fig:workspace}
\end{figure}
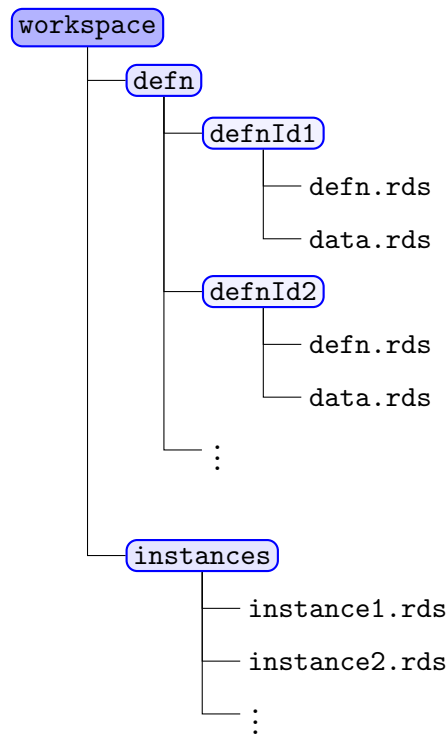

\section{User interface and further implementation details}
\label{sec:ui}
In order to make \pkg{distcomp} accessible to users who may not be
  \proglang{R} programmers, \pkg{shiny} applets are provided to aid in
  the process of defining a computation, setting up worker processes
  and in generating code for the master process.

\subsection{Define a computation}
\label{sec:define-computation}

The function \code{defineNewComputation} invokes a \pkg{shiny} web
application and Figure~\ref{fig:steps} gives an overview of what is
expected by the application. A sample screen shot is shown in
Figure~\ref{fig:shiny-define-comp}. Besides gathering some generic
information such as a name and title for the computation, a formula
for the model and a starting dataset is also taken as input. We expect
that the initial proposer has actual data on subjects from her site in
some analyzable form (CSV assumed here) with meaningful names for the
covariates. Other screens enable the user to choose from the available
computations, check data for conformability and execute a dry run. For
example, the provided formula is checked against the dataset to ensure
the model can actually be fit. The process is not complete until the
fit proceeds without error; the various buttons in the user interface
are articulated appropriately. The variables in the formula then
become the variables that \emph{all} other participating sites will
have to provide at a minimum.

An identifier (for all practical purposes unique) is generated and
associated with every computation definition. This ensures that a
worker site (i.e., the associated \proglang{R} computation engine) may
participate in more than one computation; the identifier serves to
unambiguously distinguish various computations and associate
appropriate data and functions for that computation.

The final result is a \emph{compdef} object, containing all necessary
metadata defining the computation (but no individual patient data),
that is saved in an \code{.rds} file.

\subsection{Build worker process for the computation}
\label{sec:build-worker-process}

Before a worker site can be configured for a computation, an
\pkg{opencpu} server needs to be set up and its profile modified so
that the \pkg{distcomp} package is loaded and configured with a
workspace. The workspace configuration requires a one-time editing of
a security policy (at least on the commonly used Ubuntu Linux servers)
so that the \pkg{opencpu} server may write serialized objects into the
area.

The function \code{setupWorker} invokes another \pkg{shiny} webapp that
requires several inputs: the URL of the \pkg{opencpu} server, the
\emph{compdef} (see Section~\ref{sec:define-computation}) and the site
specific data. The {shiny} webapp performs sanity checks, ensures that
the model can be fit and if everything succeeds, it uploads
information to the \pkg{opencpu} server so that the computation
becomes accessible. The site is then ready to participate in a
computation.

\subsection{Build master process for the computation}
\label{sec:build-master-process}

The function \code{setupMaster} invokes a {shiny} webapp that creates
a master process \proglang{R} object that can interact with worker
processes. Inputs to this application are the the \emph{compdef}
generated in Section~\ref{sec:define-computation} and URLs of sites
participating in the computation. Once again some sanity checks are
performed to ensure that sites are indeed addressable and finally, the
\proglang{R} code for running the computation is generated. The application
updates the user \proglang{R} profile with code that ensures the
instantiation of an object with the correct identifier and variables
and formulae.  No data is needed at this point, but the master object
can do no real computation until worker sites are added.

\subsection{Workspace Details}
\label{sec:workspace-details}

The package \pkg{distcomp} requires a workspace to do its work. The
full structure of the workspace is shown in Figure~\ref{fig:workspace}
where \proglang{R} objects are persisted for computations. The
\code{defn} folders store \emph{compdef} objects under various
identifiers generated for the computations. The \code{instances}
folders store the instantiated objects used in the computations; these
may be repeatedly saved during iterations as they change state.

A workspace is set up only once. It is best done by modifying the user
\proglang{R} profile to contain the following two lines.
\begin{CodeOutput}
library("distcomp")
distcompSetup(workspace = "full path of workspace",
              ssl.verifyhost=FALSE,
              ssl.verifypeer=FALSE)
\end{CodeOutput}
(On Unix and MacOS, the user's \code{.Rprofile} file in the home
directory will suffice, but on Windows, this needs to be inserted into
the site profile.)  Thus every \proglang{R} invocation will have
loaded the \pkg{distcomp} package and know about the workspace.

\end{document}